\documentclass[11pt,aps,prd,showpacs,nofootinbib,superscriptaddress,preprint,preprintnumbers]{revtex4}

\pdfoutput=1

\usepackage[usenames,dvipsnames]{color}
\usepackage{graphicx}
\usepackage{amsmath,amsfonts,amssymb,dsfont,bm,bbm,dcolumn,float,ifthen,mathrsfs,wasysym}
\usepackage{pstricks,feyn,verbatim,slashed,xcolor,multirow,array,rotating}
\usepackage{url,hyperref,fancybox,epstopdf}

\newcommand{\GeV}      {~\mathrm{GeV}}
\newcommand{\TeV}      {~\mathrm{TeV}}

\newcommand{\pb}      {~\mathrm{pb}}
\newcommand{\fb}      {~\mathrm{fb}}
\newcommand{\ab}      {~\mathrm{ab}}

\newcommand{\ba}{\begin{array}}
\newcommand{\ea}{\end{array}}
\newcommand{\beqn}{\begin{eqnarray}}
\newcommand{\eeqn}{\end{eqnarray}}
\newcommand{\beqs}{\begin{subequations}}
\newcommand{\eeqs}{\end{subequations}}
\newcommand{\be}{\begin{equation}}
\newcommand{\ee}{\end{equation}}
\newcommand{\non}{\nonumber \\}

\newcommand{\mathsym}[1]{{}}

\def\gU{\rm U}
\def\gSU{\rm SU}

\def\mL{\mathcal{L}}

\def\mO{\mathcal{O}}

\newcommand{\MET}{E\hspace{-0.08in}\slash_T}

\def\hf{\frac{1}{2}}

\usepackage{ulem,fancyvrb} 
\newcommand{\NC}[1]{{\color{blue} {#1}}}

\begin{document}
\title{LHC searches for heavy neutral Higgs bosons with a top jet substructure analysis}

\begin{flushright}
ADP-15-32-T934\\
\end{flushright}

\author{Ning Chen}
\email{chenning@ustc.edu.cn}
\affiliation{Department of Modern Physics, \\ University of Science and Technology of China, Hefei, Anhui, 230026, China}
\author{Jinmian Li}
\email{jinmian.li@adelaide.edu.au}
\affiliation{ARC Centre of Excellence for Particle Physics at the Terascale, Department of Physics, University of Adelaide, Adelaide, South Australia 5005, Australia}
\author{Yandong Liu}
\email{ydliu@pku.edu.cn}
\affiliation{Department of Physics and State Key Laboratory of Nuclear Physics and Technology, Peking University, Beijing 100871, China}

\begin{abstract}
We study the LHC searches for the heavy $CP$-odd Higgs boson $A$ and $CP$-even Higgs boson $H$ in the context of a general two-Higgs-doublet model. 
Specifically, we consider the decay mode of $A/H\to t \bar t $ through the $t \bar t $ associated production channels. 
In the so-called ``alignment limit'' of the two-Higgs-doublet model, this decay mode can be the most dominant one. 
By employing the HEPTopTagger and the multivariate analysis method, we present the search sensitivities for both $CP$-odd Higgs boson $A$ and $CP$-even Higgs boson $H$ via this channel with multiple top quarks at the high-luminosity LHC runs. 

\end{abstract}

\pacs{12.60.Fr, 14.80.-j, 14.80.Ec }

\maketitle


\section{Introduction}
\label{section:intro}

The discovery of the $125\,\GeV$ Higgs boson at the LHC $7\oplus 8\,\TeV$ runs manifests the Higgs mechanism~\cite{Higgs:1964ia, Higgs:1964pj, Englert:1964et, Weinberg:1967tq} and its role playing in the electroweak symmetry breaking (EWSB). 
In many of new physics models beyond the SM (BSM), the Higgs sector is extended with several scalar multiplets. 
Examples include the minimal supersymmetric extension of the SM (MSSM)~\cite{Dimopoulos:1981zb}, the left-right symmetric models~\cite{Zhang:2007da}, and the composite Higgs models~\cite{Mrazek:2011iu}. 
There are several Higgs bosons in these models, where at least one of them is identified as the $125\,\GeV$ Higgs boson discovered at LHC. 
Therefore, those extra heavy Higgs bosons are yet to be searched for by the LHC experiments running at $\sqrt{s}=(13\,\TeV\,, 14\,\TeV)$ and the future high-energy $pp$ colliders running at $\sqrt{s}=50 - 100\,\TeV$~\cite{Gomez-Ceballos:2013zzn, CEPC-SPPC-pre}.

The two-Higgs-doublet model (2HDM) was motivated to provide extra $CP$-violation sources in the scalar sector~\cite{Lee:1974jb}. 
Such a setup is also required by the MSSM, due to the cancellation of gauge anomaly. 
To search for the second Higgs doublet, it is important to find the most dominant final states that are consistent with the LHC discovery and measurements of the $125\,\GeV$ Higgs boson. 
The global fits to the $CP$-conserving 2HDM parameter space were previously carried out by several groups~\cite{Coleppa:2013dya, Craig:2013hca, Barger:2013ofa,Chowdhury:2015yja}, all of which pointed to the so-called ``alignment limit'', i.e., $c_{\beta-\alpha}\to 0$. 
In these fits, one always assumes the light $CP$-even Higgs boson $h$ in 2HDM being the $125\,\GeV$ SM-like Higgs boson discovered by LHC. 
Another 2HDM parameter to control the size of Higgs boson couplings is the ratio of two Higgs vacuum expectation values (VEVs) $t_\beta$, with the definition given in Eq.~\eqref{eq:tb}.
Obviously, it is important to focus on the parameter space that is allowed by the current global fits for the future hunting of the other heavy Higgs bosons in 2HDM. 
There have been considerable works on the LHC searches for the heavy Higgs bosons in 2HDM through different exotic decay modes, including $A\to hZ/HZ$~\cite{Coleppa:2013xfa, Coleppa:2014hxa, Dorsch:2014qja, Chen:2014dma}, $H\to hh$~\cite{Chen:2013rba, Craig:2013hca, Chen:2013emb, Baglio:2014nea, Hespel:2014sla, Barger:2014qva}, $A/H \to W^\pm H^\mp$~\cite{Li:2015lra}, $H\to H^+ H^-$~\cite{Li:2015lra}, and $H^\pm \to A W^\pm / H W^\pm$~\cite{Coleppa:2014cca}. 
The current experimental searches at the LHC $8\,\TeV$ runs include $A\to hZ$~\cite{CMS:2013eua, Aad:2015wra, Khachatryan:2015lba} and $H\to hh$~\cite{CMS:2013eua,ATLAS:2014rxa, Aad:2014yja}. 
Some of the decay modes, such as $A\to hZ$ and $H\to hh$, are due to couplings that are proportional to the alignment parameter $c_{\beta-\alpha}$. 
In the exact alignment limit of $c_{\beta - \alpha}=0$, these decay modes will be vanishing and are of minor interest for the next-step experimental searches. 
Other decay modes, such as $A\to HZ$ and $H\to H^+ H^-$, which involve another undiscovered heavy scalar in the final states, are usually suppressed by the small phase space. 
The conventional experimental searches for the heavy Higgs bosons were motivated by the MSSM scenario where the Higgs sector is define by the 2HDM-II setup. 
For the large-$t_\beta$ inputs, the heavy Higgs boson couplings with the charged leptons and the down-type quarks are enhanced. 
Correspondingly, the important experimental search modes are $A/H\to (b \bar b\,,\tau^+ \tau^-)$~\cite{Abazov:2008hu, Aaltonen:2009vf, Abazov:2011up, Chatrchyan:2012vp, Aad:2012cfr, CMS:2013hja, Chatrchyan:2013qga, Khachatryan:2014wca, Aad:2014vgg} and $H^\pm \to (t b\,, \tau^\pm\nu)$~\cite{CMS:2014pea, Aad:2014kga}, which mostly exclude the heavy Higgs bosons from the large-$t_\beta$ parameter space. 
On the other hand, the final states of $t \bar t$ from the heavy neutral Higgs boson decays can be quite important with the low- and intermediate-$t_\beta$ inputs. 
The searches for the $t \bar t$ final states from the heavy Higgs boson decays are thought to be very challenging, where one has to deal with the large SM background of $pp\to t \bar t$. 
For the $A/H$ production, it is known that the signal channel of $gg\to A/H \to t \bar t $ strongly interferes with the SM background~\cite{Dicus:1994bm, Frederix:2007gi, Jung:2015gta} and results in a peak-dip structure. 
Therefore, one can only rely on the heavy-quark associated production channels of $b \bar b + A/H$ and $t \bar t + A/H$ to study the $A/H \to t \bar t $ decays. 
The searches for processes involving these final states at the ILC, the LHC and the future $100\,\TeV$ $pp$ colliders were recently studied in Refs.~\cite{Kanemura:2014dea, Dev:2014yca,Craig:2015jba,Kanemura:2015nza, Hajer:2015gka, Kuang:2015iwa}.

In this work, we study the LHC searches for the heavy neutral Higgs bosons by tagging the boosted top jets from their hadronic decays.  
The technique of tagging boosted objects such as SM Higgs bosons~\cite{Butterworth:2008iy, Butterworth:2008tr}, top quarks~\cite{Kaplan:2008ie, Plehn:2009rk, Plehn:2010st, Plehn:2011sj, Plehn:2011tg, Plehn:2012pr, Kling:2012up, Schaetzel:2013vka, Schaetzel:2014kha, Larkoski:2014zma, Kasieczka:2015jma}, and $(W\,,Z)$ gauge bosons~\cite{Cui:2010km} were previously proposed by several different groups. 
Some of the previous studies of the BSM new state searches and the measurements of the Higgs boson properties by tagging the boosted jets include Refs.~\cite{Yang:2011jk, Yang:2014usa, Reuter:2014iya, Buckley:2015vsa}.
Generally speaking, these techniques use certain jet algorithms to reconstruct the entire hadronic decay of the boosted objects of $(H_{\rm SM}\,, t\,, W\,,Z)$, rather than reconstructing the individual decay products. 
Specifically, there are two classes of methods of tagging the boosted top quarks. 
One algorithm is called the JHUTopTagger~\cite{Kaplan:2008ie}, which requires the summation of the transverse momenta of the decayed particles to be larger than $1\,\TeV$. 
This is very challenging when one is interested in mother particles with masses of several hundred $\GeV$ to $\mO(1)\,\TeV$. 
Alternatively, we study the LHC searches for the $A/H \to t \bar t$ decays by using the HEPTopTagger method, which is efficient in tagging the top jets with intermediate transverse momenta of $\mO(100)\,\GeV$. 
We consider the $t \bar t + A/H $ production channels, with the sequential decay mode of $A/H \to t \bar t$. 
At least one hadronically decayed top quark will be reconstructed by the HEPTopTagger, and we denote such a top quark as $t_h$. 
For the $t \bar t + A/H $ channel, we look for the signal channels of $t_h$ plus the same-sign dileptons (SSDL).

Our paper is organized as follows. 
In Sec.~\ref{section:AHin2HDM}, we have a review of the heavy neutral Higgs bosons in the framework of the $CP$-conserving general 2HDM. 
By identifying the light $CP$-even Higgs boson $h$ being the $125\,\GeV$ Higgs boson discovered at the LHC, the current global fit to the Higgs signal strengths point to the alignment limit of $c_{\beta - \alpha }\to 0$ in the general 2HDM. 
In this limit, we show that the decay modes of $A/H\to t  \bar t$ are always the most dominant ones for 2HDM-I and also dominant ones for 2HDM-II with low-$t_\beta$ inputs. 
The inclusive production cross sections of $\sigma[pp\to t \bar t + (A/H \to t \bar t) ]$ at the LHC $14\,\TeV$ runs are evaluated. 
In Sec.~\ref{section:AHtott}, we perform the Monte Carlo (MC) simulations and the top tagging analysis of the $A/H \to t \bar t $ decay modes. 
We consider the heavy neutral Higgs bosons $A/H$ in the mass range of $M_{A/H}\in (350\,\GeV\,, 1200\,\GeV)$. One of the top quarks from the $A/H\to t \bar t $ decay channel will be tagged through its hadronic decay products by the HEPTopTagger method. 
For the $t \bar t + A/H$ associated production channel with the $A/H \to t \bar t $ decays, we look for signal events with a boosted top jet plus the SSDL. 
After imposing the preselections to both signals and SM background, we feed the events into the ROOT TMVA package~\cite{Hocker:2007ht} to gain more discrimination power between the signal and SM background processes. 
We obtain a signal reach for $M_{A/H}\sim \mO(1)\,\TeV $ at the $14\,\TeV$ high-luminosity LHC (HL LHC)  runs with integrated luminosities of $3\,\ab^{-1}$. 
The results are projected to the $(M_{A/H}\,, t_\beta)$ plane of the general 2HDM with the alignment limit. 
This search turns out to be most sensitive to the low-$t_\beta$ parameter space of the general 2HDM. 
Finally, our conclusion and discussion are in Sec.~\ref{section:conclusion}.


\section{The Heavy Neutral Higgs Bosons in 2HDM}
\label{section:AHin2HDM}

In this section, we briefly discuss the productions and decays of the heavy neutral Higgs bosons $A$ and $H$ in the context of the general $CP$-conserving 2HDM. 
We always take the alignment limit of $c_{\beta-\alpha}=0$ in the general 2HDM, which is consistent with the current global fit when the light $CP$-even Higgs boson $h$ is SM-like with mass of $125\,\GeV$.


\subsection{The 2HDM couplings in the alignment limit}

The most general 2HDM Higgs potential is composed of all gauge-invariant and renormalizable terms by two Higgs doublets $(\Phi_1\,, \Phi_2)\in 2_{+1}$ of the $\gSU(2)_L \times \gU(1)_Y$ electroweak gauge symmetries. 
For the $CP$-conserving case, there can be two mass terms plus seven quartic coupling terms with real parameters. 
For simplicity, we consider the soft breaking of a discrete $\mathbb{Z}_{2}$ symmetry, under which two Higgs doublets transform as $(\Phi_1\,,\Phi_2)\to (-\Phi_1\,, \Phi_2)$. 
The corresponding Lagrangian is expressed as
\beqn\label{eq:2HDM_potential}
\mL&=&\sum_{i=1\,,2}|D \Phi_i|^2 - V(\Phi_1\,,\Phi_2)\,,\\
V(\Phi_1\,,\Phi_2)&=&m_{11}^2|\Phi_1|^2+m_{22}^2|\Phi_2|^2-m_{12}^2(\Phi_1^\dag\Phi_2+H.c.)+\hf\lambda_1 |\Phi_{1}|^{4} +\hf\lambda_2|\Phi_{2}|^{4}\non
&+&\lambda_3|\Phi_1|^2 |\Phi_2|^2+\lambda_4 |\Phi_1^\dag \Phi_2|^2+\hf\lambda_5 \Big[ (\Phi_1^\dag\Phi_2) (\Phi_1^\dag\Phi_2)+H.c.\Big]\,.
\eeqn
Two Higgs doublets $\Phi_1$ and $\Phi_2$ pick up VEVs to trigger the EWSB
\beqn\label{eq:2HDM_vevs}
&&\langle \Phi_{1} \rangle =\frac{1}{\sqrt{2}}\left( \begin{array}{c} 0 \\ v_{1} \\ \end{array}  \right)\qquad  \langle \Phi_{2} \rangle =\frac{1}{\sqrt{2}}\left( \begin{array}{c} 0 \\ v_{2} \\ \end{array}  \right)\,,
\eeqn
and one parametrizes the ratio of the two Higgs VEVs as
\beqn\label{eq:tb}
t_\beta&\equiv&\tan\beta = \frac{v_2}{v_1}\,.
\eeqn
The perturbative bounds of the heavy Higgs boson Yukawa couplings constrain the choices of $t_\beta$, which should be neither as small as $\mO(0.1)$ nor as large as $\mO(50)$. 
In our discussions, we mostly focus on the parameter regions of $t_\beta \sim \mO(1)$. The light $CP$-even Higgs boson $h$ is taken as the only state in the 2HDM spectra with mass of 125 GeV and its couplings with SM fermions and gauge bosons are controlled by two parameters of $(\alpha\,,\beta)$. 
A more convenient choice of the 2HDM parameter set is $(c_{\beta-\alpha}\,, t_\beta)$. The current global fits~\cite{Coleppa:2013dya, Craig:2013hca, Barger:2013ofa,Chowdhury:2015yja} by using the LHC $7\oplus 8\,\TeV$ runs to the 2HDM parameters point to the alignment limit of $c_{\beta- \alpha}\to 0$.

\begin{table}[htb]
\begin{center}
\begin{tabular}{c|c|c}
\hline
 & 2HDM-I & 2HDM-II    \\\hline\hline 
 $\xi_H^{u}$  & $-1/t_{\beta}$ & $-1/t_{\beta}$  \\
 $\xi_H^{d}$  & $-1/t_{\beta}$ & $t_{\beta}$  \\
 $\xi_H^{\ell}$  & $-1/t_{\beta}$ & $t_{\beta}$ \\
 \hline 
 $\xi_{A}^{u}$  & $1/t_{\beta}$ & $1/t_{\beta}$  \\
 $\xi_{A}^{d}$  & $-1/t_{\beta}$ & $t_{\beta}$  \\
 $\xi_{A}^{\ell}$  & $-1/t_{\beta}$ & $t_{\beta}$ \\
 \hline
\end{tabular}
\caption{The Yukawa couplings of the SM quarks and charged leptons to the heavy $CP$-even Higgs boson $H$ and the $CP$-odd Higgs boson $A$ in 2HDM-I and 2HDM-II with the alignment limit of $c_{\beta-\alpha}=0$.}\label{tab:yuk}
\end{center}
\end{table}

In the general 2HDM, SM fermions with the same quantum numbers couple to the same Higgs doublet, which will avoid the tree-level flavor-changing neutral currents. 
For 2HDM-I, all SM fermions couple to one Higgs doublet (conventionally chosen to be $\Phi_2$). This setup can be achieved by assigning a discrete $\mathbb{Z}_2$ symmetry under which $\Phi_1 \to - \Phi_1$. 
For 2HDM-II, the up-type quarks $u_i$ couple to one Higgs doublet (conventionally chosen to be $\Phi_2$) and the down-type quarks $d_i$ and the charged leptons $\ell_i$ couple to the other ($\Phi_1$). 
This can also be achieved by assigning a discrete $\mathbb{Z}_2$ symmetry under which $\Phi_1 \to - \Phi_1$ together with $(d_i\,, \ell_i)\to (- d_i \,, - \ell_i)$. 
Details of the 2HDM Yukawa setups were reviewed in Ref.~\cite{Branco:2011iw}. 
At the tree level, the heavy $CP$-even Higgs boson $H$ and the $CP$-odd Higgs boson $A$ couple to the SM fermions through the Yukawa coupling terms
\beqn\label{eq:Yukawa}
-\mL_Y&=&\sum_f  \frac{m_f}{v} ( \xi_H^f  \bar f f H -i\xi_A^f \bar f \gamma_5 f A)\,,
\eeqn
where $f$ is the SM fermion, $m_f$ is the SM fermion mass, and $v=\sqrt{v_1^2 + v_2^2}= (\sqrt{2}\, G_F)^{-1/2}=246\,\GeV$. 
The Higgs Yukawa couplings are simplified greatly when one sets in the alignment limit of $c_{\beta-\alpha}=0$. 
The dimensionless coupling strengths of $\xi_H^f$ and $\xi_A^f$ in the alignment limit are listed in Table.~\ref{tab:yuk} for the 2HDM-I and 2HDM-II cases, which only depend on $t_\beta$. 
In both 2HDM-I and 2HDM-II cases, the dimensionless Yukawa couplings of $\xi_{A/H}^u$ are inversely proportional to $t_\beta$. 
Therefore, the cross sections of $t \bar t + (A/H \to t \bar t) $ are always enhanced in the low-$t_\beta$ regions.

Besides the Yukawa couplings, there are also other couplings relevant to the decays of the heavy neutral Higgs bosons $A$ and $H$. 
From the 2HDM kinematic terms $|D \Phi_i |^2$, one has Higgs-gauge couplings of $G(HVV)$, $G(AhZ/AHZ)$, and $G(A H^\pm W^\mp)$. 
The $G(HVV)$ and $G(AhZ)$ couplings are vanishing in the $c_{\beta - \alpha}=0$ limit. 
From the general 2HDM potential, one also has the triple Higgs couplings such as $G(Hhh)$, $G(HAA)$, and $G(H H^+ H^-)$. 
The existences of these couplings lead to exotic heavy Higgs boson search strategies.

Though we focus on the alignment limit of $c_{\beta-\alpha}=0$, it is noted that the current global fits to the $125\,\GeV$ $CP$-even Higgs boson $h$ in 2HDM generally allow the parameter choices of $c_{\beta-\alpha}\sim \mO(0.1)$ for 2HDM-I and $c_{\beta-\alpha}\sim \mO(0.01)$ for 2HDM-II, respectively. 
The relevant results were obtained in Refs.~\cite{Coleppa:2013dya, Craig:2013hca, Barger:2013ofa,Chowdhury:2015yja}. 
It was shown in the previous discussions~\cite{Chen:2014dma, Chen:2013emb} that some of the heavy Higgs boson decay modes of $A\to hZ$ and $H\to hh$ can become the leading ones, especially for $M_{A/H}\lesssim 2 m_t$. 
The relevant LHC searches can be performed by the boosted Higgs searches plus the opposite-sign-same-flavor di-leptons~\cite{Chen:2014dma}, and via the $b \bar b+ \gamma\gamma$ final states~\cite{Chen:2013emb}. 
Note that the enhancements of such decay modes may rely on some special choices of parameters. 
For example, the $A\to hZ$ can be significant when masses of two $CP$-even states are nearly degenerate, and the \NC{$H\to hh$ mode is} mostly important when the corresponding Higgs cubic couplings in the 2HDM potential are enhanced. 
In this sense, the full discussions of parameter regions deviating from the alignment limit rely not only on the values of $c_{\beta-\alpha}$ but also on the input parameters of the full 2HDM mass spectrum and the 2HDM potentials as well. 
Later, we present our search results in terms of the model-independent cross sections and map into the $(M_{A/H}\,,t_\beta)$ plane for 2HDM-I and 2HDM-II under the alignment limit. 
The results can be rescaled to other general parameter inputs straightforwardly. 
As we show below, the decay modes of $A/H\to t \bar t$ are particularly relevant for the low-$t_\beta$ parameter regions of 2HDM, which was also pointed out in Ref.~\cite{Djouadi:2015jea} for the MSSM case.


\subsection{The heavy quark associated productions and decays of $A$ and $H$}

In the alignment limit of $c_{\beta -\alpha }=0$, the production channels of the heavy neutral Higgs bosons $A/H$ at the LHC include the gluon-gluon fusion and the heavy quark associated processes~\cite{Djouadi:2005gi, Djouadi:2005gj}. 
The previous studies of the gluon-gluon fusion to $t \bar t$ final states via the spin-0 resonances have suggested strong interference effects with the QCD backgrounds. 
Therefore, the heavy-quark associated productions are considered as the production channels at the LHC. 
For the 2HDM-I case, all dimensionless Yukawa couplings of the SM fermions to the $A/H$ are universally proportional to $1/t_\beta$. 
One expects the scaling of the production cross sections $\sigma [pp\to b \bar b + A/H]\sim 1/t_\beta^2$ and $\sigma [pp \to  t \bar t + A/H]\sim 1/t_\beta^2 $ with various $t_\beta$ inputs. 
For the 2HDM-II case, the dimensionless Yukawa couplings scale as $\xi_{A/H}^u\propto 1/t_\beta$ and $\xi_{A/H}^d \propto t_\beta$, respectively. 
Therefore, the contributions from the top quark annihilation processes will be enhanced significantly with the small-$t_\beta$ inputs.
Note that for heavy Higgs bosons in the mass range of $M_{A/H}\in (350\,\GeV\,, 1200\,\GeV)$, the total decay width of $\Gamma_{A/H}$ can be as large as $\mO(1)-\mO(10)\,\GeV$ with the small-$t_\beta$ inputs. Consequently, there can be potentially large interference effects between the signals and the QCD background. 
Through the evaluations by MadGraph5$\_$aMC@NLO~\cite{Alwall:2014hca}, it turns out that the inclusion of the interference effects leads to $\sim\mO(10\,\%)$ corrections to the signal cross sections.

 \begin{table}[h]
\begin{center}
\begin{tabular}{|cclc|c} \hline    
 $A$ decays & Final states &   & Alignment limit \\
 \hline\hline
 \multirow{2}{*}{SM fermions}  & $A\to (\tau^+ \tau^- \,, \mu^+ \mu^-) $  & & $\checkmark$  \\
 & $A\to (t \bar t\,, b \bar b ) $  & & $\checkmark$  \\
 \hline
 \multirow{3}{*}{Exotics}  & $A\to h Z$ &   & $-$ \\
 & $A\to H Z$ &  & $\checkmark$  \\
 & $A\to H^\pm W^\mp$ &   & $\checkmark$ \\
 \hline
 Loops & $A\to( gg\,, \gamma \gamma \,, \gamma Z)$  & & $\checkmark$ \\
\hline
\end{tabular}
\caption{The classification of the $CP$-odd Higgs boson $A$ decay modes in the general 2HDM. A checkmark (dash) indicates that the decay mode is present (absent) in the $c_{\beta-\alpha}=0$ alignment limit.  }\label{tab:Adecay_align}
\end{center}
\end{table}

\begin{table}[h]
\begin{center}
\begin{tabular}{|cclc|c} \hline    
 $H$ decays & Final states &   & Alignment limit \\
 \hline\hline
 \multirow{2}{*}{SM fermions} & $H\to (\tau^+ \tau^- \,, \mu^+ \mu^-) $  & & $\checkmark$  \\
  & $H\to (t \bar t\,, b \bar b) $  & & $\checkmark$  \\
 \hline
gauge bosons & $H\to (WW\,, ZZ) $  & & $-$  \\
 \hline
\multirow{5}{*}{Exotics} & $H\to AZ$ &   & $\checkmark$ \\
 & $H\to H^\pm W^\mp$ &   & $\checkmark$ \\
 & $H\to h h$ &   & $-$ \\
 & $H\to AA$ &   & $-$ \\
 & $H\to H^+ H^-$ &  & $\checkmark$ \\
 \hline
Loops & $H\to(gg\,, \gamma \gamma \,, \gamma Z)$  & & $\checkmark$ \\
\hline
\end{tabular}
\caption{The classification of the $CP$-even Higgs boson $H$ decay modes in the general 2HDM. A checkmark (dash) indicates that the decay mode is present (absent) in the $c_{\beta-\alpha}=0$ alignment limit. }\label{tab:Hdecay_align}
\end{center}
\end{table}

\begin{figure}
\centering
\includegraphics[width=6.5cm,height=4.5cm]{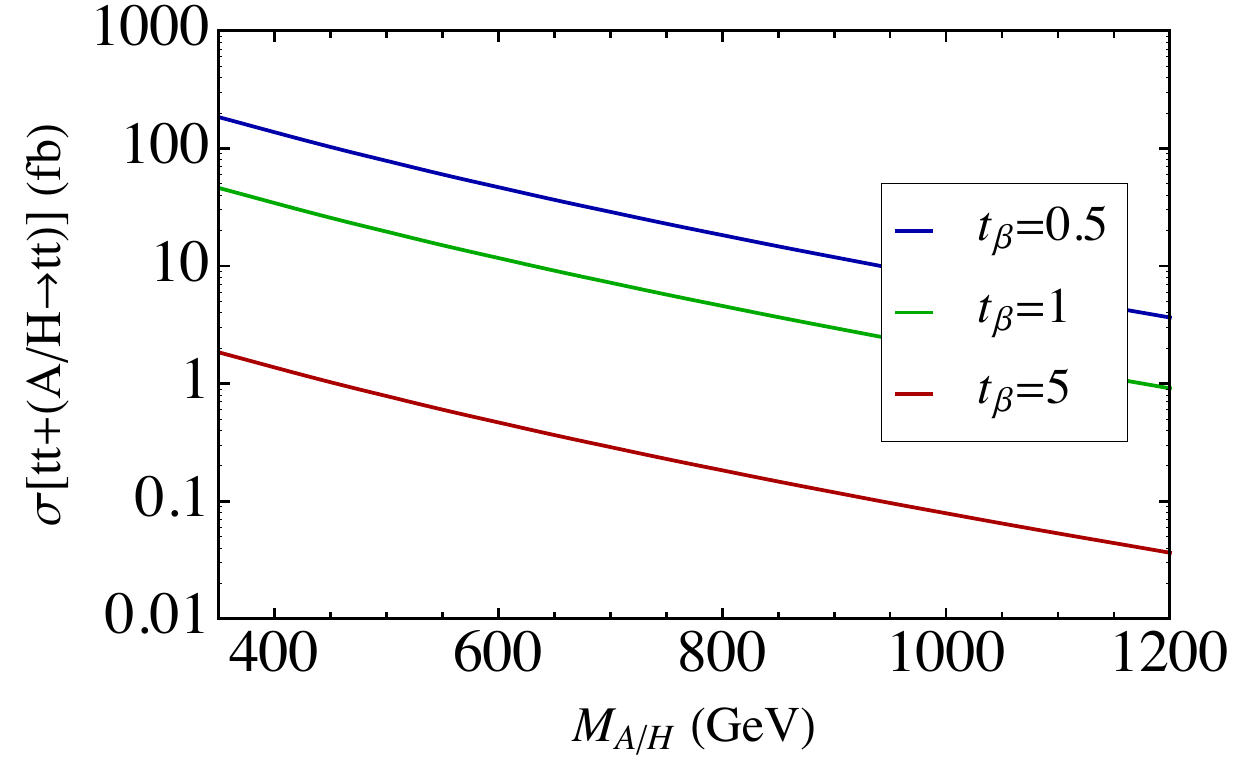}
\caption{\label{fig:AHttxsec} Shown is $\sigma[pp\to t \bar t A/H]\times {\rm BR}[A/H\to t \bar  t]$ (for both 2HDM-I and 2HDM-II cases) with $M_{A/H}\in (350\,\GeV, 1200\,\GeV)$ at the LHC $14\,\TeV$ runs.}
\end{figure}

All possible decay modes of heavy neutral Higgs boson are listed in Tables.~\ref{tab:Adecay_align} and \ref{tab:Hdecay_align} for $A$ and $H$, respectively. 
For our discussions, we evaluate their partial decay widths in the mass range of $M_{A/H}\in (350\,\GeV\,, 1200 \,\GeV)$ by 2HDMC~\cite{Eriksson:2009ws}. 
Besides of setting the alignment limit of $c_{\beta - \alpha } = 0$, we also turn off all possible exotic decay modes listed in Tables.~\ref{tab:Adecay_align} and \ref{tab:Hdecay_align}. 
This is reasonable when one assumes the masses of all heavy Higgs bosons are close to each other: $M_A\sim M_H \sim M_{H^\pm}$. 
Therefore, the only relevant decay modes for heavy neutral Higgs bosons are $A/H \to \bar f f$ and $A/H \to gg$. 
The loop-induced decay branching ratios of ${\rm Br}[A/H \to \gamma\gamma/ \gamma Z]$ are typically smaller than $10^{-5}$, which can be neglected. 
For the 2HDM-I case, the decay branching ratios of ${\rm Br}[A/H\to t \bar  t]$ are always dominant to be $\sim \mO(1)$ since all dimensionless Higgs Yukawa couplings scale as $\sim 1/t_\beta$. 
For the 2HDM-II case, the ${\rm Br}[A/H\to t \bar  t]$ can be suppressed to $\sim\mO(0.1)$ with the large-$t_\beta$ inputs, where the partial decay widths of $\Gamma [A/H \to b \bar  b]$ and $\Gamma[A/H \to \tau^+ \tau^-]$ become dominant. 
Combining with the production cross sections evaluated by MadGraph5$\_$aMC@NLO previously, we demonstrated the cross sections of $\sigma[pp\to t \bar t+A/H] \times {\rm Br}[A/H\to t \bar t  ]$ within the mass range of $M_{A/H}\in (350\,\GeV\,, 1200\,\GeV)$ at the LHC $14\,\TeV$ runs in Fig.~\ref{fig:AHttxsec}. 
As it turns out, the decay branching ratios of ${\rm Br}[A/H \to t \bar t]$ tend to unity for both 2HDM-I and 2HDM-II with the small-$t_\beta$ inputs of $\sim \mO(1)$. 
For this reason, we combine the cross sections of $\sigma[pp\to t \bar t + A/H] \times {\rm Br}[A/H\to t \bar t  ]$ for both 2HDM-I and 2HDM-II into one plot.


\section{The LHC Searches for The Heavy Neutral Higgs Bosons via The $t  \bar t$ Channel}
\label{section:AHtott}

In this section, we analyze the LHC $14\,\TeV$ searches for the heavy neutral Higgs bosons $A$ and $H$ via the $t \bar t + A/H$ productions, with the sequential decay modes of $A/H\to t \bar t $. 
We always tag the boosted top jets $t_h$ by using the HEPTopTagger method. 
Afterwards, we shall look for events including a top jet $t_h$ plus SSDL. 
The corresponding SM background processes should include the final states with SSDL plus multiple jets, where a jet may be mistagged as the boosted $t_h$. 
Thus, the SM backgrounds include $t \bar t $~\cite{Ahrens:2011px, Czakon:2013goa}, $t \bar t  b \bar b$, $(W^\pm Z\,, ZZ)$ plus jets~\cite{Campbell:2011bn}, and $(t \bar t W^\pm\,, t \bar t Z)$~\cite{kang:2014jia}. 
Here, we evaluate all relevant cross sections of the SM background processes by MadGraph5$\_$aMC@NLO~\cite{Alwall:2014hca} with the NLO QCD corrections
\beqn\label{eq:SM_bkg}
&&\sigma(pp\to t \bar t)\approx 803\,\pb\,,\non
&& \sigma (pp \to  t \bar t b \bar b )\approx 30\,\pb\,, \non
&& \sigma (pp \to  W^\pm Z )\approx 50 \,\pb\,, \non
&& \sigma (pp \to  ZZ )\approx 15 \,\pb\,.
\eeqn
To evaluate the signal significance, we should take into account the uncertainties from the SM background processes.
For dominant SM backgrounds of $t \bar t$ and $W^\pm Z$, the naive estimation of the signal significance may be made by using the theoretical uncertainties of $(\delta \sigma/\sigma)_{t \bar t}\approx 10\,\%$~\cite{Ahrens:2011px, Czakon:2013goa} and $(\delta \sigma/\sigma)_{ W^\pm Z }\approx 5\,\%$~\cite{Campbell:2011bn} due to the factorization scale uncertainties.
Furthermore, the MC uncertainties at the extremely constrained regions of phase space can be typically larger. 
To account for the MC uncertainties, we use the uncertainties of $\sim 20\,\%$ for the $t\bar t$ process and the uncertainties of $\sim 10\,\%$ for the $W^\pm Z$ as the conservative estimations of the signal reaches.
There are other SM background processes including $(W^\pm W^\pm\,, t \bar t + W^\pm\,, t \bar t + Z)$ plus jets, with the corresponding cross sections less than $1\,\pb$. 
As we show later, the dominant SM background processes after the preselections of $t_h$ plus SSDL are $t \bar t$ and $W^\pm Z$ plus jets. 
Therefore, we neglect all other SM background processes in our later analysis.
After the reconstruction of the boosted $t_h$, we shall select the kinematic variables for the signal and background events and carry out the TMVA analysis to optimize the rate of $S/\sqrt{B}$. 
The results of our analysis will be projected to the signal reaches at the HL LHC runs with integrated luminosities of $3\,\ab^{-1}$.

\subsection{The MC simulations and the top jet tagging}

For event generations of the signal processes, we use Universal FeynRules Output~\cite{Christensen:2008py} simplified models with $A$ or $H$ being the only BSM particles. 
The relevant coupling terms to be implemented are the Yukawa couplings of $A t \bar t /Ht \bar t$. 
We generate events for both signal and SM background processes at the parton level by MadGraph5$\_$aMC@NLO~\cite{Alwall:2014hca}, with the subsequent parton shower and hadronization performed by Pythia~\cite{Sjostrand:2006za}. 
Afterwards, Delphes~\cite{deFavereau:2013fsa} is used for the fast detector simulations. 
In our simulations of both signal and background processes, we include up to two extra jets with the MLM matching in order to avoid the double counting. 
Our fast detector simulations follow the setup of the ATLAS detector.

\begin{table}[htb]
\begin{center}
\begin{tabular}{c|c|c|c|c|c}
\hline
$M_{A/H}$ & $350\,\GeV$ & $400\,\GeV$  & $500\,\GeV$ & $600\,\GeV$ & $700\,\GeV$    \\\hline
$R_{\rm CA}$ & $2.3$   & $2.2$   & $2.1$ & $ 2.1$ & $ 2.0$ \\\hline\hline
$M_{A/H}$ & $800\,\GeV$   & $900\,\GeV$ & $1000\,\GeV$ & $1100\,\GeV$  & $1200\,\GeV$    \\\hline
$R_{\rm CA}$ & $2.0$ & $1.9$ & $1.8$ & $1.8$ & $1.7$ \\
 \hline
\end{tabular}
\caption{ The choices of the jet cone sizes $R_{\rm CA}$ for different $M_{A/H}$ inputs via the $t \bar t + A/H$ (second entries) production channels. }\label{tab:Ropt}
\end{center}
\end{table}

\begin{figure}
\centering
\includegraphics[width=6.5cm,height=4.5cm]{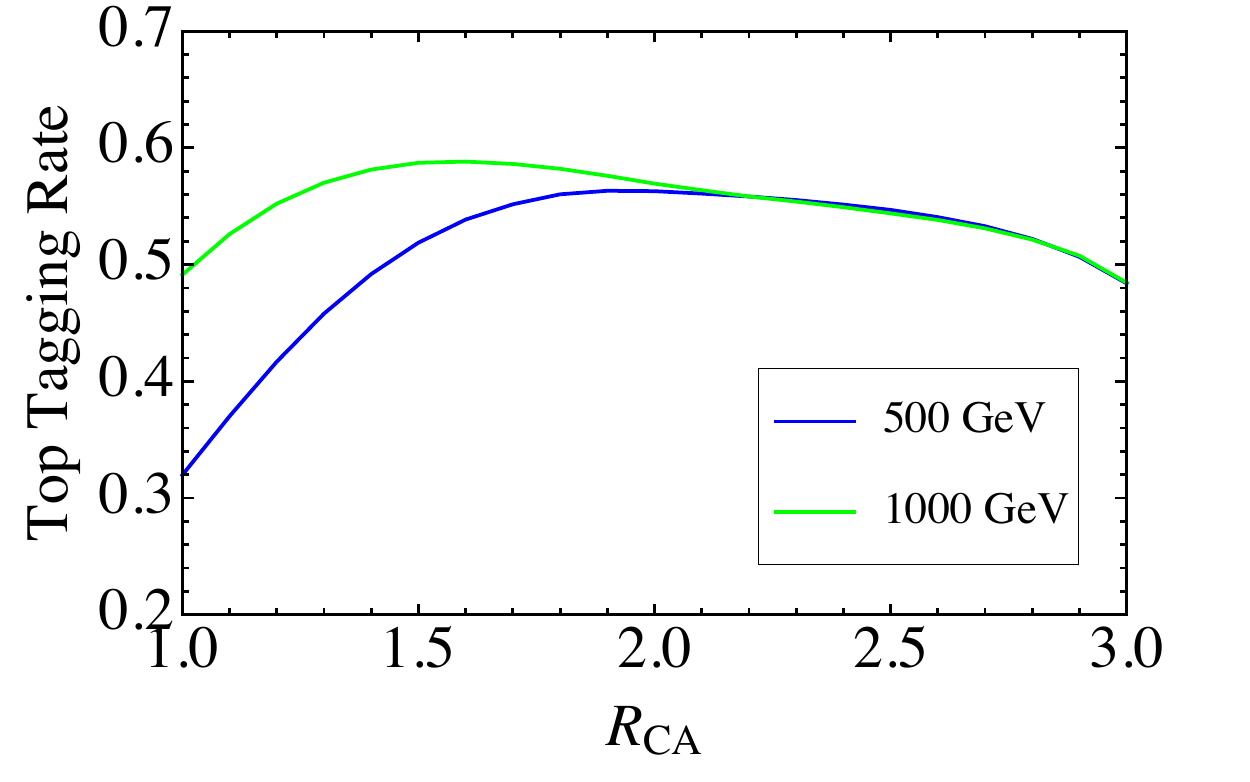}
\caption{\label{fig:TopRate} The tagging rates of the top jet $t_h$ versus the CA jet cone sizes $R_{\rm CA}\in (1.0\,, 3.0)$. Two samples of $M_{A/H}=500\,\GeV$ (blue curve) and $M_{A/H}=1000\,\GeV$ (green curve) are shown. }
\end{figure}

In what follows, we briefly describe the reconstruction of physical objects by the HEPTopTagger method. 
The details of the algorithm can be found in the original Refs.~\cite{Plehn:2009rk, Plehn:2010st, Plehn:2011sj}. 
The energy flow observables from the Delphes output are used for the jet substructure analysis by Fastjet~\cite{Cacciari:2011ma}. 
In each event, we cluster the top jets by using the Cambridge-Aachen (CA) algorithm~\cite{Dokshitzer:1997in, Wobisch:1998wt} with certain jet cone size $R_{\rm CA}$. 
By setting the reconstructed top mass range of $ m_t^{\rm rec} \in (140\,\GeV \,, 210\,\GeV)$, the HEPTopTagger algorithm finds a candidate boosted top jet which contains three subjets with their total transverse momenta greater than $200\,\GeV$. 
The rate of tagging one $t_h$ can be $\sim 30\% - 60\%$ with certain choices of $R_{\rm CA}$. As an illustration, we plot the tagging rates of the boosted top jet $t_h$ versus the CA jet cone sizes for the $M_{A/H}=500\,\GeV$ and $M_{A/H}=1\,\TeV$ samples in Fig.~\ref{fig:TopRate}. 
It is also likely to tag two boosted top jets at the rate of $\sim 10\% - 20\%$. For such cases, we always choose the one with the largest $p_T$ as the $t_h$. 
Generally speaking, the tagging rates of top jets vary with different choices of the jet cone sizes $R_{\rm CA}$. The boost factors of top jets are enhanced with the heavier resonances of $M_{A/H}$. 
For each mass of signal process $pp \to t \bar t + A/H $, we scan the jet cone sizes $R_{\rm CA}\in (1.0\,, 3.0)$ at the step of $0.1$ for reconstructing the top jet $t_h$ in the HEPTopTagger. 
In addition, the effects due to the underlying events can be eliminated by the filtering procedure~\cite{Butterworth:2008iy} in the HEPTopTagger.
In Table.~\ref{tab:Ropt}, we list our choices of $R_{\rm CA}$ for the sample models with masses of $M_{A/H}\in (350\,\GeV\,, 1200\,\GeV)$. 
The selection criteria is given below for two different production channels. Once a $t_h$ is obtained in an event, we erase its constituents from the input particles. 
The remaining particles will be clustered into narrow jets by using the anti-$k_t$ algorithm with a jet cone size of $R_{\rm narrow}=0.4$. 
The narrow jets are required to satisfy $p_T \ge 20\,\GeV $ and $|\eta|<4.5$.

\subsection{Multivariable analysis }

\begin{figure}
\centering
\includegraphics[width=6cm,height=4cm]{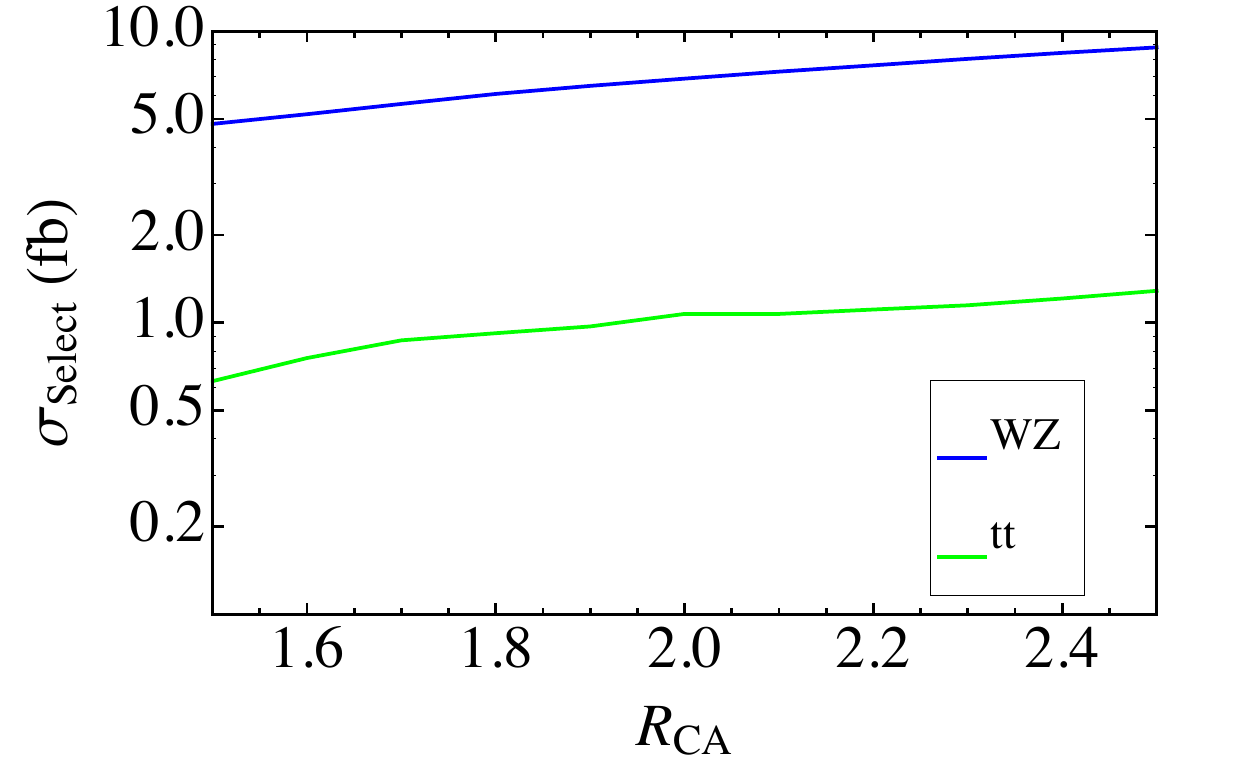}
\caption{\label{fig:ttAbkg_selection} The SM background cross sections after the $t_h$ plus SSDL preselections versus the CA algorithm jet cone sizes. }
\end{figure}

For the $t \bar t + (A/H \to t \bar t)$ signal channel, one has four top quarks in the final states. 
As stated previously, at least one top quark $t_h$ from the $A/H$ decay should be reconstructed by the HEPTopTagger method through its hadronic decay mode. 
Afterwards, we select events containing SSDL $\ell_1^\pm \ell_2^\pm$ from the semi-leptonic decays of two other top quarks. 
It turns out a significant suppression to the SM background can be achieved by selecting the events containing $t_h$ plus SSDL. 
An example of the preselection efficiencies of events for the $M_{A/H}=500\,\GeV$ case is tabulated in Table.~\ref{tab:MA500eff_14TeV}. 
The suppression rates of SM background events from the $t \bar t$ and $t \bar t b \bar b$ can be as significant as $\sim 10^{-5}$ when imposing the SSDL selection criterion. 
Obviously, the $W^\pm Z$ background becomes the most dominant one after the preselections. 
Meanwhile, one has $\sigma(t \bar t)_{\rm select} \approx 0.1\, \sigma(W^\pm Z)_{\rm select} $ after the preselections. 
In Fig.~\ref{fig:ttAbkg_selection}, we display the cross sections of $W^\pm Z$ and $t \bar t $ after selecting the $t_h$ plus the SSDL events versus different choices of the jet cone sizes $R_{\rm CA}$.

\begin{table}[t]
\begin{center}
\begin{tabular}{c|c|ccccccc}
\hline
 $M_{A/H}$ & Signal & $t \bar t $ & $t \bar t  b \bar b $ & $W^\pm Z$ & $ZZ$  & $S/\sqrt{B}$    \\\hline\hline  
 Total cross section $({\rm fb})$ & $50$ & $8.0\times 10^5$ & $3\times 10^4$ & $5.0\times 10^4$ & $1.5\times 10^4$  & $...$  \\
 Preselection of $t_h $ $({\rm fb})$ & $28$ & $3.1\times 10^5$ & $1.4\times 10^4$ & $7.7 \times 10^3$ & $1.9 \times 10^3$  &  $...$ \\
Preselection of $t_h + $SSDL $({\rm fb})$ & $0.48$ & $0.56$ & $0.11$ & $3.92$ & $0.17$  &  $2.3\, (1.2)$ \\
\hline
 \end{tabular}
\caption{
The preselection efficiencies of the $M_A=500\,\GeV$ (with $R_{\rm CA}=2.1$) signal and background processes at the $14\,\TeV$ HL LHC. 
We assume the nominal cross section for the signal process to be $\sigma[pp\to t \bar t + A/H]\times {\rm BR}[A/H\to t \bar t ]=50\,\fb$. 
The signal significances are obtained by the naive estimation (the conservative estimation) of the SM background uncertainties.
}\label{tab:MA500eff_14TeV}
\end{center}
\end{table}

\begin{figure}
\centering
\includegraphics[width=6.5cm,height=4.5cm]{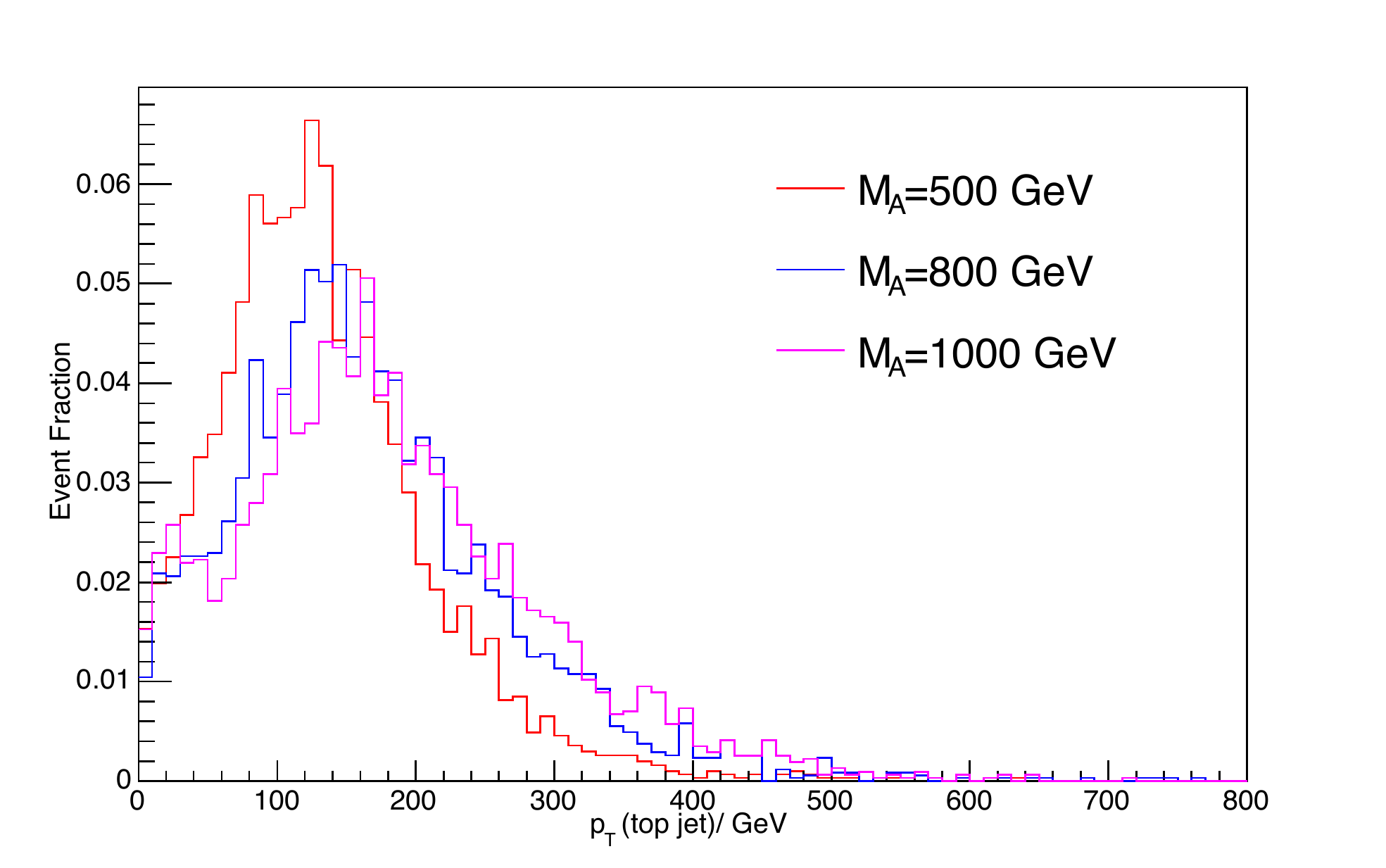}
\includegraphics[width=6.5cm,height=4.5cm]{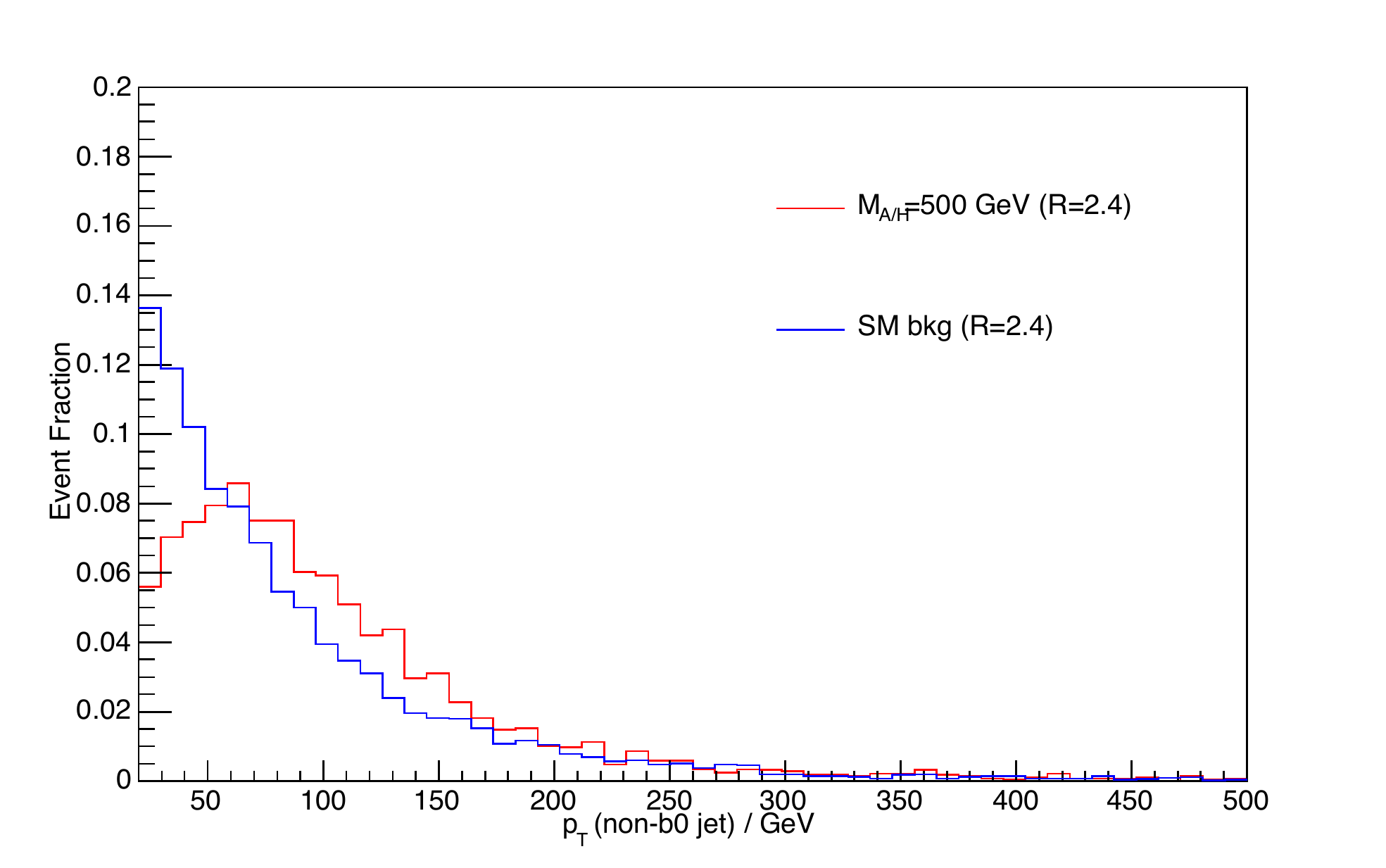}
\includegraphics[width=6cm,height=4cm]{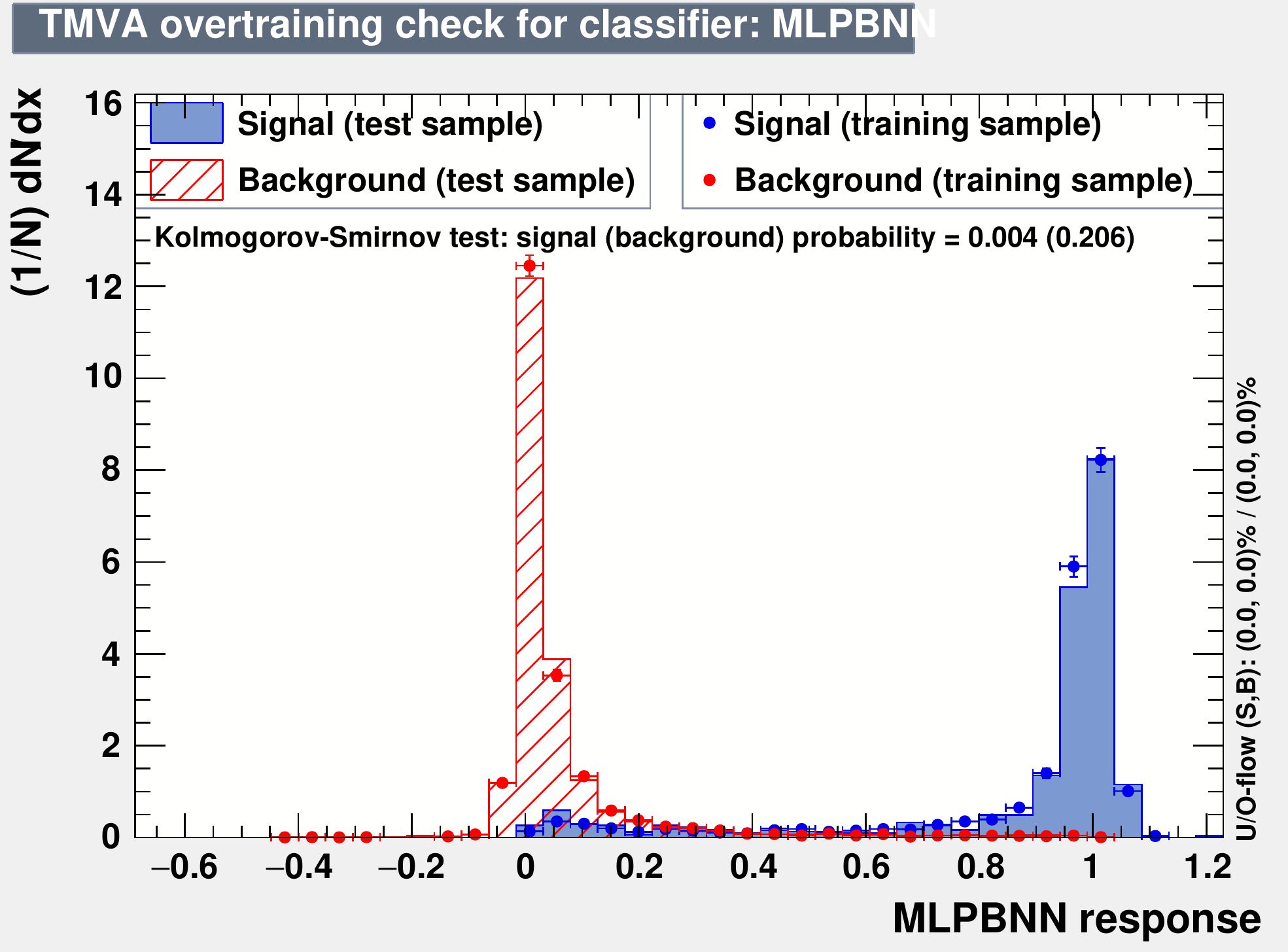}
\includegraphics[width=6cm,height=4cm]{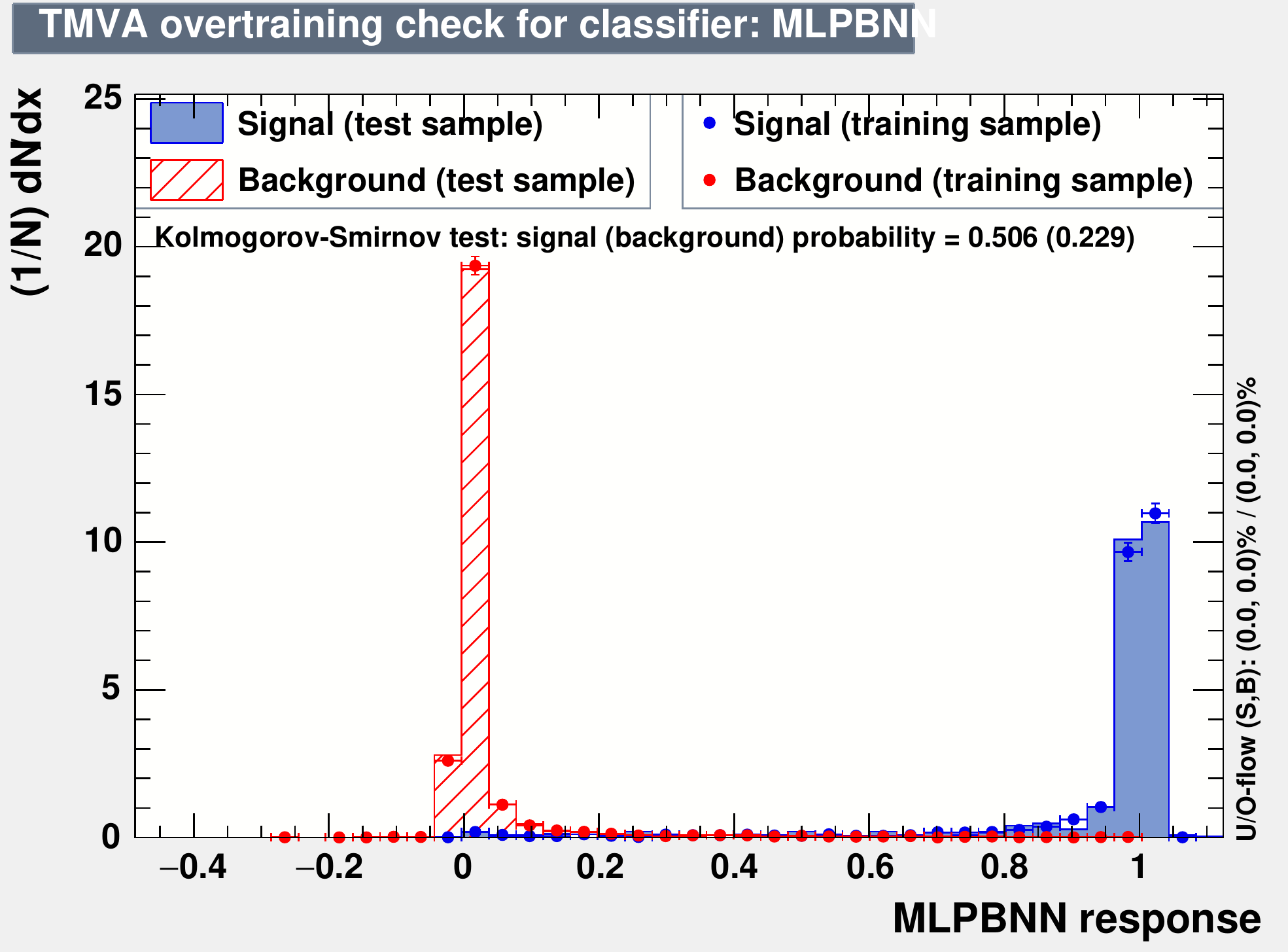}
\caption{\label{fig:ttAH_kin} Upper: The $p_T$ of the $t_h$ for the signal processes $t \bar t + (A/H\to  t \bar  t)$ with $M_{A/H}= (500\,\GeV\,, 800\,\GeV\,, 1000\,\GeV )$ (left panel) and the $p_T$ of the leading narrow jet reconstructed by the anti-$k_t$ algorithm for $M_{A/H}= 500\,\GeV$ (right panel) and SM background processes after the preselections. Lower: The normalized distributions of MLP neural network response for signal and background for the $t \bar t + (A/H \to t \bar t )$ channel, left: $M_{A/H}=500 \,\GeV$, right: $M_{A/H}=1000 \,\GeV$.}
\end{figure}

\begin{figure}
\centering
\includegraphics[width=6.5cm,height=4.5cm]{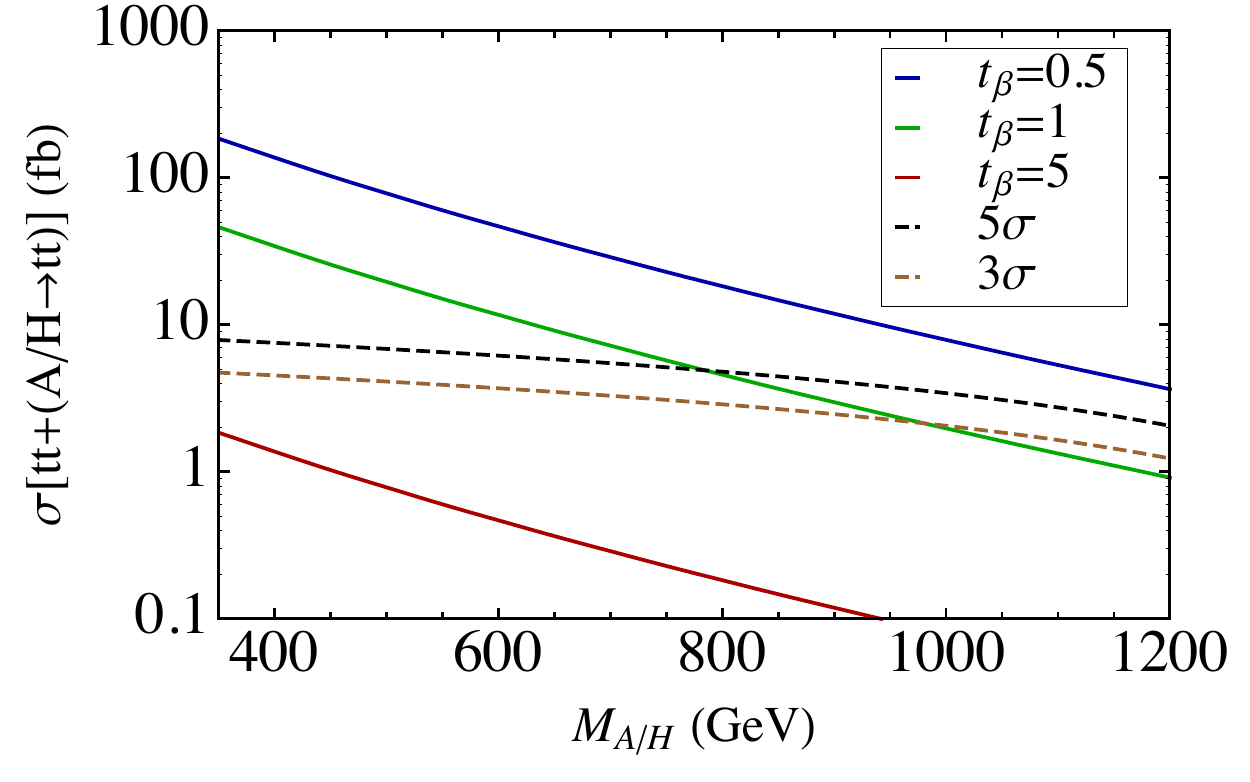}
\includegraphics[width=6.5cm,height=4.5cm]{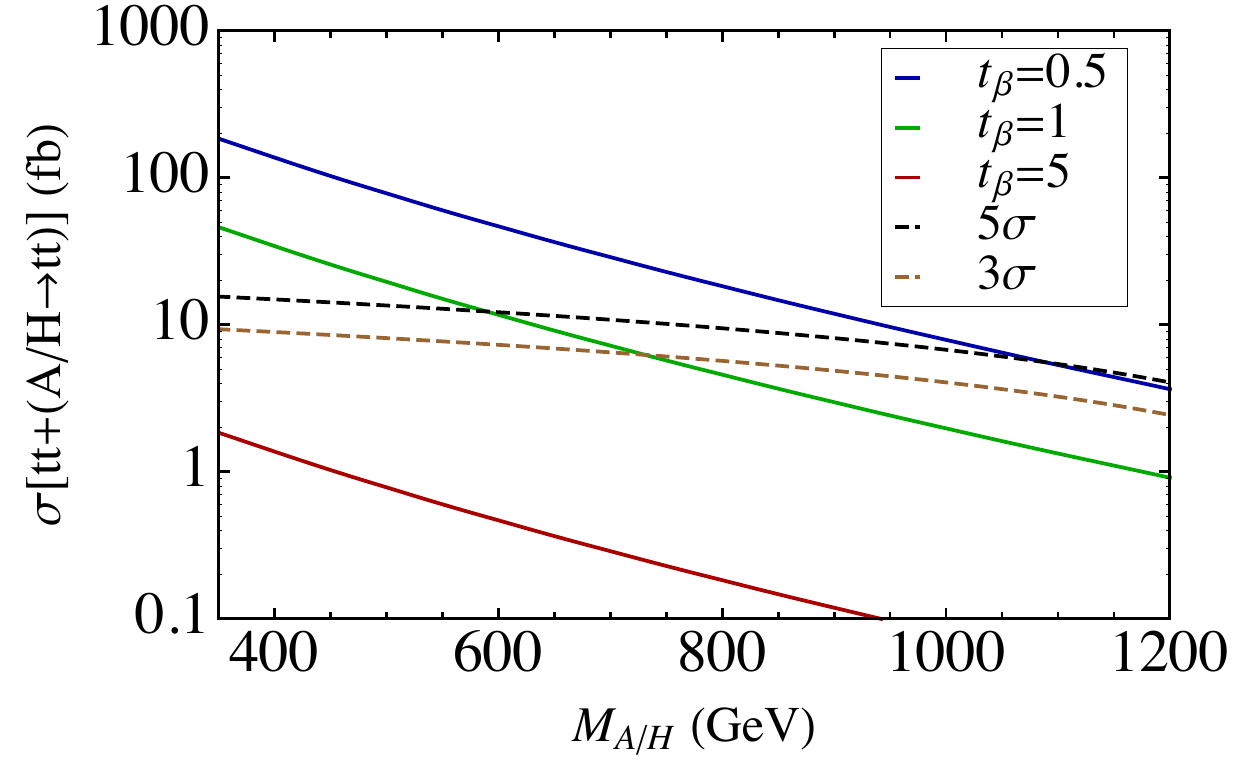}
\includegraphics[width=6.5cm,height=4.5cm]{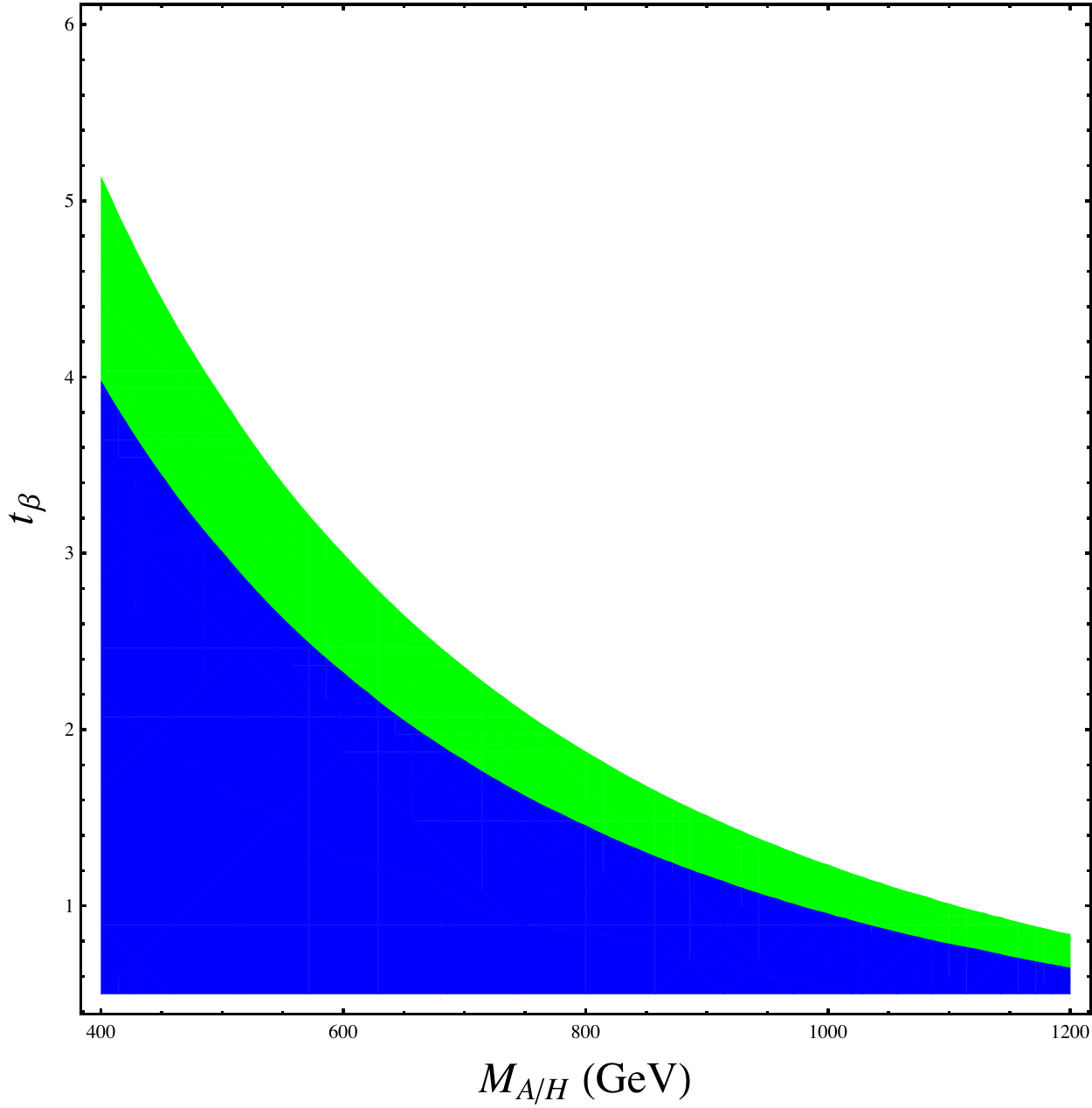}
\includegraphics[width=6.5cm,height=4.5cm]{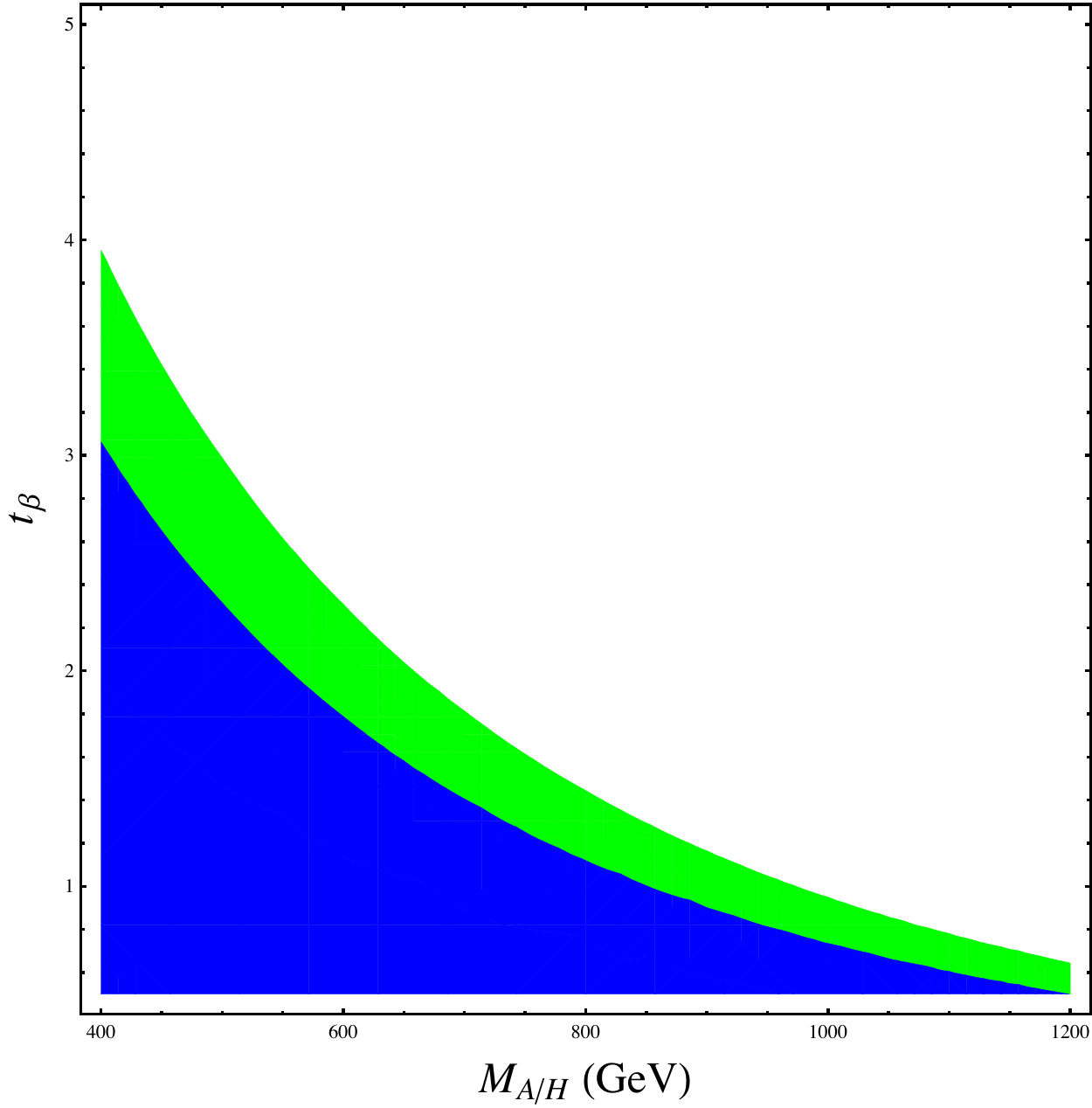}
\caption{\label{fig:ttAHreach} 
The signal predictions together with the signal reaches (dashed lines) of $t \bar t + (A/H \to t \bar t )$ at the HL LHC runs. 
Upper: The mode-independent cross section reaches at the HL LHC runs.
Lower: The $5\,\sigma$ (in blue) and $3\,\sigma$ (in green) signal reaches projected on the $(M_{A/H}\,, t_\beta)$ plane at the HL LHC runs.
We present results by using both naive estimations (left panels) and conservative estimations (right panels) of the SM background uncertainties. 
}
\end{figure}

To further discriminate the signal events from the SM background, we adopt the MLP neural network analysis in the ROOT TMVA package. 
The list of kinematic variables for the multivariate analysis include $(p_T\,, \eta\,, \phi)$ of $(t_h\,, \ell_1^\pm\,, \ell_2^\pm)$, $\MET$, number of $( b\textrm{-jets}\,, \textrm{ non-$b$ jets})$, $p_T(b_0\,, j_0)$, $\sum_j p_T(j)$, and $\sum_b p_T(b)$. 
Here $j_0$ and $b_0$ denote the leading non-$b$-jet and the leading $b$-jet ordered in their transverse momenta, respectively. 
A $b$-tagging rate of 70 \%~\cite{ATLAS:2012aoa} is assumed to reconstruct the $b$-jets.
For signal events with different masses of $M_{A/H}$, the $p_T$ distributions of the tagged $t_h$ are shown in the upper left panel of Fig.~\ref{fig:ttAH_kin}. 
For the samples shown here, the $p_T$ of the tagged $t_h$ is not a very significant discriminator among different signal samples. 
Due to this reason, we find the signal reaches for different $M_{A/H}$ inputs are pretty close within the mass range of the study. 
On the upper right panel of Fig.~\ref{fig:ttAH_kin}, we display the distributions of the $p_T(j_0)$ between the signal processes versus the SM backgrounds for the $M_{A/H}=500\,\GeV$ signal sample. 
There is clear discrimination between the signal and background events on the $p_T$ of narrow jets. 
Afterwards, we feed the events passing the preselections into the TMVA package for further optimization of the cuts among all kinematic variables listed above. 
As it turns out, the kinematic variables of $p_T(b_0\,, j_0)$ and the number of $b\textrm{-jets}$ are the leading important ones for the TMVA to apply cuts. 
The discriminations between signal and SM backgrounds from the TMVA analysis are also presented for the $M_{A/H}= 500\,\GeV$ and $M_{A/H}=1000\,\GeV$ samples. 
After obtaining the cut efficiencies, we convert the results to the signal cross sections within the $5\,\sigma$ discovery limits, which read $\sigma[  t \bar t + (A/H \to t \bar t ) ] \sim 5-8\,\fb$ (naive estimation) or $\sigma[  t \bar t + (A/H \to t \bar t ) ] \sim 10-15\,\fb$ (conservative estimation) with $M_{A/H}\in (350\,, 1200)\,\GeV$. 
The results are demonstrated in the left panel of Fig.~\ref{fig:ttAHreach}. 
By looking for the $t_h$ plus the SSDL signals, our analysis shows that the HL LHC searches are likely to reach the heavy neutral Higgs boson masses up to $\mO(1)\,\TeV$ in the low-$t_\beta$ regions for the general $CP$-conserving 2HDM. 
The model-independent signal cross sections for the $5\,\sigma$ reaches are further projected to the $(M_{A/H}\,, t_\beta)$ plane, as shown in the right panel of Fig.~\ref{fig:ttAHreach}. 
The previous Ref.~\cite{Kanemura:2015nza} studied the inclusive searches for the four top final states from the 2HDM heavy Higgs bosons at the LHC, where they obtained a mass reaches of $\mO(800)\,\GeV$ with $t_\beta \lesssim 1.5$ at the $2\,\sigma$ C.L.\,. 
As a comparison, our search strategies are likely to improve the signal reaches for four top final states with mass up to $\mO(1)\,\TeV$ at the $5\,\sigma$ C.L. thanks to the reconstruction of boosted top jets and the TMVA analysis.


\section{Conclusion and Discussion}
\label{section:conclusion}

In this paper, we have carried out an analysis of the heavy neutral Higgs boson searches via the $t \bar  t$ decay modes at the LHC $14\,\TeV$ runs. 
In the simplified scenario where one sets the alignment limit of $c_{\beta-\alpha}=0$ and turns off all possible exotic decay modes of heavy neutral Higgs bosons, the decay branching ratios of $t \bar t $ final states can be usually approaching $\mO(1)$ with low-$t_\beta$ inputs. 
Correspondingly, the searches for the $A/H \to t \bar t$ are of the top priority from the perspective of the production cross sections. 
We consider the $t \bar t + (A/H \to t \bar t)$ signal channels in this work, whose interference effects with the QCD background are less severe compared to the gluon fusion channel. 
In order to suppress the corresponding SM background contributions, we adopt the HEPTopTagger method to tag the boosted top jets $t_h$ for both signal processes. 
Because the boost factor thus the spreading angle of $t_h$ is mainly controlled by the mass of $A/H$, for each $M_{A/H}$ input, the cone size parameter of the HEPTopTagger is optimized to gain the maximal signal top tagging rate.
For the $t \bar t  + (A/H \to t \bar t)$ signal channel, we require a boosted $t_h$ plus the SSDL in the final state.
Afterwards, the MLP neural network analysis is applied based on the TMVA package. 
For $M_{A/H}\in (350\,\GeV\,, 1200\,\GeV)$, the production cross sections of $t \bar t + (A/H \to t \bar t)$ as small as $\sim [5-8]\,\fb $ (naive estimation) or $\sim [10-15]\, \fb$ (conservative estimation) can be discovered at $5\,\sigma$ C.L.. 
Based on these results, we eventually obtain the LHC signal reaches for the heavy neutral Higgs bosons via the $t \bar t $ final states on the $(M_{A/H}\,, t_\beta)$ plane. Our results are most sensitive to the heavy neutral Higgs boson searches with the low-$t_\beta$ inputs, which are complementary to the previous LHC searches performed for the $( b \bar b\,, \tau^+ \tau^-)$ final states. For the 2HDM inputs of $t_\beta\sim \mO(1)$ with the alignment limit, the $M_{A/H}\lesssim 1\,\TeV$ region can be reached at the HL LHC runs via the boosted top jet tagging plus the SSDL final state searches.


\section*{Acknowledgments}

We thank Jan Hajer, Tao Liu, Yanwen Liu, Michael Spannowsky and Qishu Yan for very helpful discussions. We also thank Minghui Liu and Haiping Peng for setting up the computation environment for our work. This work is partially supported by National Science Foundation of China (under Grants No. 11275009, No. 11335007, No. 11575176), the Fundamental Research Funds for the Central Universities (under Grant No. WK2030040069), and the Australian Research Council (under Grant No. CE110001004). N.C. would like to thank the Institute for Modern Physics of Zhejiang University and the Center of High Energy Physics (CHEP) of Peking University for their hospitality when part of this work was prepared.

\end{document}